\documentclass{appolb}
\usepackage[top=3cm, bottom=3cm, left=3cm, right=3cm]{geometry}
\usepackage{amsmath}
\usepackage{amsfonts}
\usepackage{amssymb}
\usepackage{graphicx}
\graphicspath{{FiguresPDF/}}
\usepackage[colorlinks,citecolor=blue,linktoc=all,linkcolor=cyan]{hyperref}
\usepackage{graphicx}
\usepackage{svg}
\usepackage{mathtools}
\usepackage{upgreek}
\usepackage{physics}
\usepackage{multirow}
\usepackage{placeins} 
\usepackage{pgfplots}
\usepackage{subfigure}
\usepackage{caption}
\usepackage{array}

\usepackage[T1]{fontenc}
\usepackage{dsfont}               
\usepackage{mathrsfs}             
\usepackage{slashed}              
\usepackage{amsbsy}
\usepackage{braket}
\usepackage{tikz}
\usepackage{bbold}

\renewcommand{\thetable}{\Roman{table}}

\providecommand{\keywords}[1]
{	
  \textbf{\textit{Keywords---}} #1
}

\usepackage[
    backend=biber,
    citestyle=numeric-comp,
    bibstyle=science,
    sorting=none,
    eprint=true,
    url=false,
    doi=true]{biblatex}
\DeclareFieldFormat{eprint:arxiv}{%
  Arxiv:\space
  \href{https://arxiv.org/abs/#1}{#1}%
  \iffieldundef{eprintclass}{}{ \mkbibbrackets{\thefield{eprintclass}}}%
}

\renewbibmacro*{doi+eprint+url}{%
  \printfield{doi}%
  \iffieldundef{doi}{%
    \iffieldundef{journaltitle}{%
      \setunit{\addspace}%
      \printfield[eprint:arxiv]{eprint}%
    }{}%
  }{}%
}

\DeclareSourcemap{
  \maps[datatype=bibtex]{
    \map{
      \step[fieldset=title, null]
    }
  }
}

\usepackage{xpatch}
\xpatchbibmacro{name:andothers}{ 
{\finalandcomma}%
}{%
\addspace%
}{}{}

\DeclareFieldFormat{doi}{%
  \mkbibacro{DOI}\addcolon\space
  \ifhyperref
    {\href{https://doi.org/#1}{\nolinkurl{#1}}}
    {\nolinkurl{#1}}}

\DeclareFieldFormat{labelnumberwidth}{\mkbibbrackets{#1}}

\AtEveryBibitem{\clearfield{month}}

\providecommand{\U}[1]{\protect\rule{.1in}{.1in}}

\pgfplotsset{compat=1.8}

\usepackage{authblk}          

\title{From breakfast, lunch, and dinner to the Bell-inequality tetrahedron }

\author[1,2]{Francesco Giacosa\thanks{francesco.giacosa@ujk.edu.pl}}

\affil[1]{Institute of Physics, Jan Kochanowski University, Kielce, Poland}
\affil[2]{Institute for Theoretical Physics, J. W. Goethe University, Frankfurt am Main, Germany}

\begin{document}

\sloppy 
\maketitle

\begin{abstract}
A simple pedagogical discussion and visualization of certain Bell inequalities
is presented with the help of a system involving two \textquotedblleft
twins\textquotedblright, originally discussed in 2605.03104. These twins answer
identically when the same `question' is posed, here exemplified with:  Did you like breakfast/lunch/dinner? In analogy, for two
entangled particles at distant locations, the `twin-like' quantum state is such that the
same outcome for the same measurement (i.e. spin direction) is obtained. In
the space $(X,Y,Z)$ of mixed moments (i.e. $X$ refers to breakfast-lunch correlation, being $1$ for same answers and $-1$ for opposite
ones), the space of Bell local models is contained in a tetrahedron, while the
quantum space is bound by an elliptope that embeds the tetrahedron. The whole
no-signalling region in the  $(X,Y,Z)$ space is the cube $[-1,1]^{3}.$ The twin scenario applies e.g. to the decay of the Higgs into fermions.

\end{abstract}

\bigskip

Bell inequalities show which part of the probability space is possible for
classical local models \cite{Bell:1964kc,Bell:1980wg,Gisin:2007gps}. 
The probability space refers to all possible outcomes
when a given number of observers perform, at their own location, a certain
measurement which can result in different outcomes. 
A full classification of Bell inequalities is mathematically quite complicated
and reveals a rich structure that generally results in polytopes, where the realization
of each local model is bound to lie in its interior or at most at its
boundary \cite{Bancal:2010vhq,Brunner:2013est,Rosset:2014tsa}. On the other hand, Quantum Mechanics (QM) allows for a quantum region in probability space that
contains the local space but is typically striclty larger: there are regions which are possible for QM, but are
not such for local models. The physical realization of such points represents
an experimental violation of the Bell inequalities, hence showing that local models cannot
describe Nature. This is indeed what various experiments show, e.g. Refs. \cite{Aspect:1982,Hensen:2015,Giustina:2015}.

Even further, one recognizes no-signalling space, that is all those models
that do not allow for signaling: the outcome probability at each site does
not depend on the measurement performed at distant locations \cite{Masanes:2005njk}. All in all, the
hierarchy is:
\[
\mathcal{L} \subsetneq \mathcal{Q} \subsetneq  \mathcal{NS}~,
\]
where $\mathcal{L}$ stands for the local, $\mathcal{Q}$ for the quantum, and $\mathcal{NS}$ for the no-signalling regions, respectively \cite{Brunner:2013est}.  

Here, based on the recent work of Ref. \cite{Gazdzicki:2026grg}, we discuss a specific simple realization
and visualization of $\mathcal{L}$, $\mathcal{Q}$, and $\mathcal{NS}$. Our presentation is pedagogical
and as free as possible from technicalities. In Ref. \cite{Gazdzicki:2026grg}, a so-called symmetric local Bell setup was studied in the
framework of three different measurement types (also called colloquially as
questions, identical for each observer) $q_{k=1,2,3}$ and two separate
observers ($1 \equiv$ Alice and $2 \equiv$ Bob), whose outcomes, also denoted as answers to the questions above, are
of the type $a_{i=1,2}=\pm1$ (yes/no). With symmetric Bell, it is intended
that, for any given value of a suitable hidden variable $\lambda$ that describes shared resources of the two observers, the probability
outcomes at both sites are equal:
\begin{equation}
p_{1}(\pm1|q_{k},\lambda)=p_{2}(\pm1|q_{k},\lambda)=p(\pm1|q_{k}%
,\lambda)\text{ }\forall k=1,2,3\text{ .}%
\label{stoctwin}
\end{equation}
Moreover, we impose the usual Bell factorization
\begin{equation}
p(a_{1},a_{2}|q_{1},q_{2},\lambda)=p(a_{1}|q_{1},\lambda)p(a_{2}|q_{2}%
,\lambda)\text{ .}%
\end{equation}
For instance, the quantity $p(1,-1|2,1,\lambda_{0})$ is the probability that, for a given hidden
variable $\lambda_{0},$ Alice answers `yes' to the question
$k=2$ and Bob `no' to $k=1$.
The final probability reads
\begin{equation}
p(a_{1},a_{2}|q_{1},q_{2})=\int d\lambda f(\lambda) p(a_{1}|q_{1},\lambda)p(a_{2}|q_{2}%
,\lambda)\text{,}%
\end{equation}
where $f(\lambda)$ is the probability distribution of the hidden variable $\lambda$ (with $\int d\lambda f(\lambda) =1$).
Note that $f(\lambda)$ does not depend on the questions or measurement settings. Sometimes this property is called free will, but measurement independence is more appropriate. Removing this assumption may lead to superdeterminism \cite{tHooft:2007oio,Hall:2010zzf,Hall:2015lil,Hossenfelder:2019shy}. 

Within this set-up, we concentrate on averages/correlators of the types
\begin{equation}
\left\langle a_{1}a_{2}\right\rangle_{q_{1}q_{2}}\text{ ,}%
\end{equation}
with special attention on the mixed moments
\begin{equation}
X=\left\langle a_{1}a_{2}\right\rangle _{12}\text{ , }Y=\left\langle
a_{1}a_{2}\right\rangle _{13}\text{ , }Z=\left\langle a_{1}a_{2}\right\rangle
_{23}%
\end{equation}
Specifically, $X$ is the correlator of the answers of both observers,
when Alice is asked $k=1$ and Bob $k=2$ (or vice versa). Its explicit form reads: 
\begin{equation}
 X   =\sum_{a_{1},a_{2}=-1}^{1}a_{1}a_{2}p(a_{1},a_{2}|1,2) = p(1,1|0,1)+p(-1,-1|1,2)-p(1,-1|1,2)-p(-1,1|1,2) \text{ ,}
\end{equation}
while $Y$ and $Z$ are obtained by replacing $p(a_{1},a_{2}|1,3)$ and $p(a_{1},a_{2}|2,3)$, respectively. 

It turns out that, in the $XYZ$ space, each symmetric
local model must lie within $\mathcal{L} \equiv$)tetrahedron\footnote{This tetrahedron differs from that of Ref.~\cite{Tavakoli:2020mkb}:
there the tetrahedral structure refers to the measurement settings, whereas
here it describes the allowed correlations $(X,Y,Z)$.} (see below for an
independent derivation). The corresponding quantum system, also fulfilling
the symmetric requirement, is contained within an elliptope $\mathcal{Q}$
that embeds the $\mathcal{L}$ region. Both are contained into the
no-signalling $\mathcal{NS}$ cube $[-1,1]^{3}$.

In this work, we give a simple derivation of this tetrahedron.
To this end, instead of `stochastic' twins as in Eq. (\ref{stoctwin}), we consider
`deterministic twins': the answer at each run is identical if the same
question is asked. We regard Alice and Bob as `perfect' twins. As such, they
like the very same food. We envisage the following experiment. Alice and Bob
are kept separate, and each day they receive breakfast, lunch, and dinner. In
the end of each day, Alice is asked one question, e.g. `did you like lunch
today?'; the answers can be $a_{1}=\pm1$ (1 for like, $-1$ for dislike), where
$k=1,2,3$ refers to lunch, breakfast, and dinner respectively. The same applies
for Bob, but the question asked to him may differ from Alice's. Thus, in a given day, a possible chart of answers can be as follows.

\begin{center}
$%
\begin{tabular}
[c]{||l||l||l||}\hline\hline
Question & Answer Alice (1) , $a_{1}$ & Answer Bob (2), $a_{2}$\\\hline\hline
$\theta_{1}$: Did you like breakfast? & yes , 1 & yes , 1\\\hline\hline
$\theta_{2}$: Did you like lunch? & yes , 1 & yes , 1\\\hline\hline
$\theta_{3}$: Did you like dinner? & no , -1 & no , -1\\\hline\hline
\end{tabular}
$
\end{center}
For this particular situation: $x=1,$ $y=-1,$ $z=-1$, thus
$x+y+z=-1\geq-1.$ Here, small $x,y,z$ refer to a specific outcome (single run), while $X,Y,Z$ refer to their average over many events/questions, for instance:
\begin{equation}
X=\lim_{N\rightarrow\infty}\frac{x_1+...+x_N}{N} \text{ ,}
\end{equation}
where $x_k$ with $k=1,2,..$ refers to all cases when Alice is asked about breakfast and Bob about lunch (or vice versa); then, all these specific outcomes are averaged. 

Another possible set of answers can be:
\begin{center}
$%
\begin{tabular}
[c]{||l||l||l||}\hline\hline
Question & Answer Alice , $a_{1}$ & Answer Bob , $a_{2}$\\\hline\hline
$\theta_{1}$: Did you like breakfast? & yes , 1 & yes , 1\\\hline\hline
$\theta_{2}$: Did you like lunch? & yes , 1 & yes , 1\\\hline\hline
$\theta_{3}$: Did you like dinner? & yes , 1 & yes , 1\\\hline\hline
\end{tabular}
$ 
\end{center}
leading to $x=y=z=1.$ Here, $x+y+z=3\geq-1$.

The whole chart of possible answers is listed below, each of them being associated with a probability $C_k$.
\begin{center}%
\begin{tabular}
[c]{||c||c||c||c||c||c||c||}\hline\hline
Breakfast & Lunch & Dinner & $x$ & $y$ & $z$ & Probability\\\hline\hline
+1 & +1 & +1 & +1 & +1 & +1 & $C_{1}$\\\hline\hline
+1 & +1 & -1 & +1 & -1 & -1 & $C_{2}$\\\hline\hline
+1 & -1 & +1 & -1 & +1 & -1 & $C_{3}$\\\hline\hline
+1 & -1 & -1 & -1 & -1 & +1 & $C_{4}$\\\hline\hline
-1 & +1 & +1 & -1 & -1 & +1 & $C_{5}$\\\hline\hline
-1 & +1 & -1 & -1 & +1 & -1 & $C_{6}$\\\hline\hline
-1 & -1 & +1 & +1 & -1 & -1 & $C_{7}$\\\hline\hline
-1 & -1 & -1 & +1 & +1 & +1 & $C_{8}$\\\hline\hline
&  &  &  &  &  & $\sum_{k=1}^{8}C_{k}=1$\\\hline\hline
\end{tabular}
\end{center}

\begin{figure}[htb]
\centering
\includegraphics[width=0.72\textwidth]{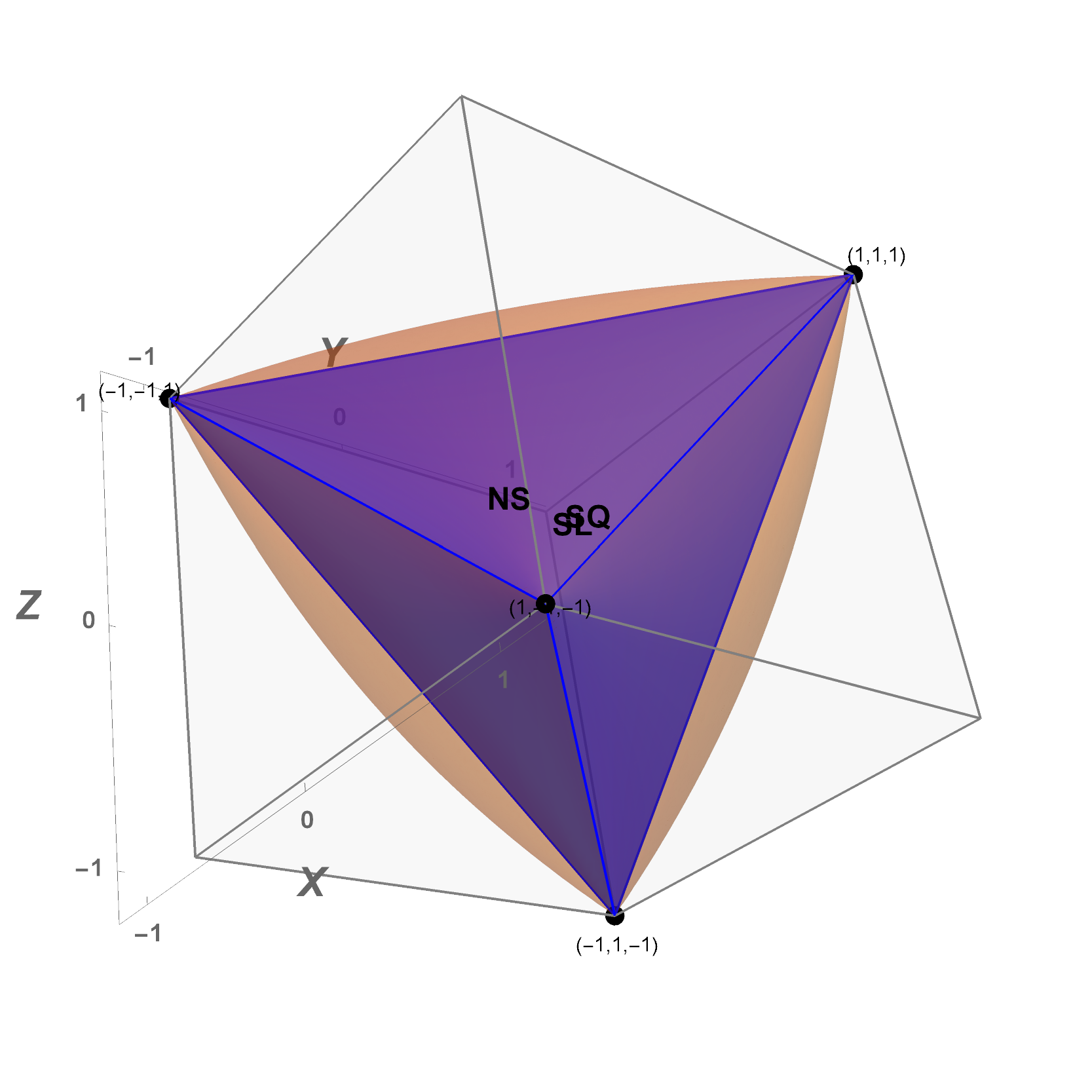}
\caption{Geometric representation of the symmetric Bell scenario in the
$(X,Y,Z)$ correlation space. The symmetric local region $\mathcal{L}$
is the tetrahedron with vertices $(1,1,1)$, $(1,-1,-1)$,
$(-1,1,-1)$, and $(-1,-1,1)$. The symmetric quantum region
$\mathcal{Q}$ is the elliptope defined by
$1+2XYZ-X^{2}-Y^{2}-Z^{2}\geq0$, which contains the local
tetrahedron. Both regions are contained in the no-signalling region
$\mathcal{NS}$, represented by the cube $[-1,1]^3$. The viewing angle has been chosen to appreciate the quantum elliptope covering the local region. }
\label{fig:tetrahedron}
\end{figure}

\begin{figure}[htb]
\centering
\includegraphics[width=\textwidth]{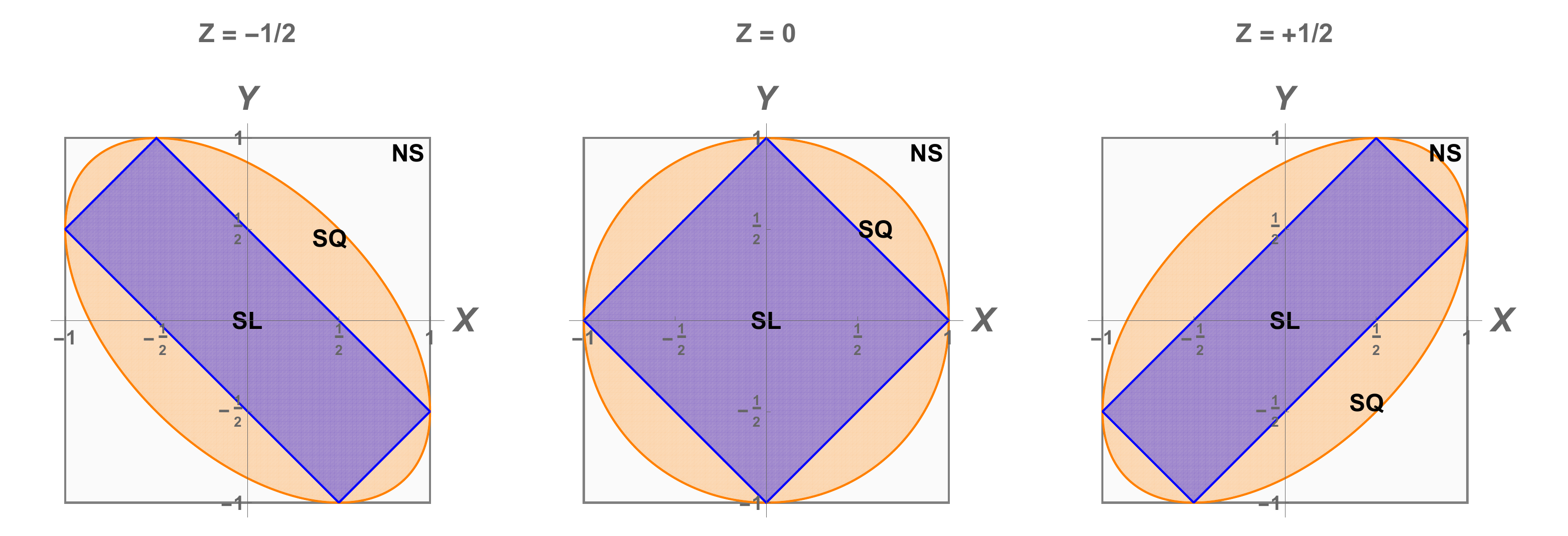}
\caption{Sections of the geometric regions of Fig.~\ref{fig:tetrahedron}
in the $(X,Y)$ plane for fixed values $Z=-1/2$, $Z=0$, and $Z=+1/2$
(from left to right). The blue region is the symmetric local region
$\mathcal{L}$, the orange region together with the blue region forms
the symmetric quantum region $\mathcal{Q}$, and the full square
corresponds to the no-signalling region $\mathcal{NS}$. For $Z=0$,
the quantum boundary reduces to the unit circle
$X^{2}+Y^{2}=1$. For $Z=\pm1/2$, the quantum sections are tilted
ellipses.}
\label{fig:sections}
\end{figure}

Every row corresponds to one of four $(x,y,z)$ points: $(1,1,1),$ $(1,-1,-1),$ $(-1,1,-1),$ and $(-1,-1,1).$ Four general
inequalities that define the tetrahedron with these vertices follow:
\begin{align}
X+Y+Z &  \geqslant-1\text{ , }X-Y+Z\leq1 \text{ ,}\\
X+Y-Z &  \leq1\text{ , }X-Y-Z\geq-1 \text{ .}
\end{align}
Namely, each of these inequalities is fulfilled for any single entry, so it is also fulfilled when averaged. 
More formally, the moments $X,Y,Z$ are given by:
\begin{align}
X  &  =C_{1}+C_{2}-C_{3}-C_{4}-C_{5}-C_{6}+C_{7}+C_{8}\\
Y  &  =C_{1}-C_{2}+C_{3}-C_{4}-C_{5}+C_{6}-C_{7}+C_{8}\\
Z  &  =C_{1}-C_{2}-C_{3}+C_{4}+C_{5}-C_{6}-C_{7}+C_{8}%
\end{align}
Then, for example:
\begin{align}
X+Y+Z  &  =3C_{1}-C_{2}-C_{3}-C_{4}-C_{5}-C_{6}-C_{7}+3C_{8}=3(C_{1}+C_{8}%
)-\sum_{k=3}^{7}C_{k}= \nonumber \\
&  =-1+4(C_{1}+C_{8})\geq-1 \text{ .}
\end{align}
A similar strategy leads to the formal proof of the other inequalities. Fig. \ref{fig:tetrahedron} shows the tetrahedron, and Fig. \ref{fig:sections} shows its sections.

Two comments are in order.

1) It is instructive to check how the usual CHSH inequality \cite{CHSH1969} behaves in the
present symmetric `twin' setting. The CHSH combination (for two questions per site only) reduces to
\begin{equation}
C_{\rm CHSH}=\left\langle a_{1}a_{2}\right\rangle _{11}+\left\langle a_{1}a_{2}\right\rangle _{12}+\left\langle a_{1}a_{2}\right\rangle _{21}-\left\langle a_{1}a_{2}\right\rangle _{22}=2X ,
\end{equation}
because $\left\langle a_{1}a_{2}\right\rangle _{11} = \left\langle a_{1}a_{2}\right\rangle _{22} = 1$. 
Since $|X|\leq 1$, the CHSH bound $|C_{\rm CHSH}|\leq2$ is then
always fulfilled. This example shows that CHSH needs different questions/settings per site to be nontrivial. 

2) It is worth discussing the relation between the exact twins used in
the present pedagogical construction and the stochastic twins considered in
Ref.~\cite{Gazdzicki:2026grg}. The former satisfy the stronger, event-by-event
condition that Alice and Bob give the same answer whenever the same question
is posed. The stochastic condition of Eq. (\ref{stoctwin})
requires only equality of the corresponding local response probabilities.
Although these two notions of twins are different at the microscopic level,
they lead, after projection onto the three mixed moments $(X,Y,Z)$, to the
same local tetrahedron. Thus, the simple deterministic construction presented
here provides an elementary realization of the same reduced local geometry
obtained from the more general stochastic symmetric model.

What about the quantum model analogous to the classical scenario outlined
above? This is a bipartite system described by a statistical operator
$\hat{\rho}$ and by three measurement settings per site, fulfilling the
following requirements for the quantum probabilities:
\begin{align}
p_{Q}(a_{1},a_{2}|q_{1},q_{2})
&=
p_{Q}(a_{2},a_{1}|q_{2},q_{1}),
\\
p_{Q}(1,1|q,q)+p_{Q}(-1,-1|q,q)
&=1.
\label{quantumtwin}
\end{align}
The first condition expresses exchange symmetry, whereas the second is the
exact-twin condition: whenever the same question is posed at the two sites,
the two answers coincide with probability one. In the literature, this quantum setup is also called synchronous \cite{Russell:2020xho,Dykema_2015}.

For dichotomic observables with outcomes $a_i=\pm1$, the quantum probability
reads (e.g. Ref. \cite{Barr:2024djo})
\begin{equation}
p_{Q}(a_{1},a_{2}|q_{1},q_{2})
=
\operatorname{Tr}\left[
\hat{\rho}
\left(
\frac{1+a_{1}\hat{O}^{(1)}_{q_{1}}}{2}
\otimes
\frac{1+a_{2}\hat{O}^{(2)}_{q_{2}}}{2}
\right)
\right],
\end{equation}
where $\hat{O}^{(1)}_{q_{1}}$ and $\hat{O}^{(2)}_{q_{2}}$ are the
dichotomic observables associated with the questions/settings $q_{1}$ and
$q_{2}$ at the two sites. The corresponding operators
$(1+a_i\hat{O}^{(i)}_{q_i})/2$ are the projectors onto the outcomes
$a_i=\pm1$.

The quantum mixed moment $X$ reads
\begin{equation}
X
=
\sum_{a_{1},a_{2}=\pm1}
a_{1}a_{2}
p_{Q}(a_{1},a_{2}|1,2)
=
\operatorname{Tr}\left[
\hat{\rho}
\left(
\hat{O}^{(1)}_{1}\otimes\hat{O}^{(2)}_{2}
\right)
\right],
\end{equation}
and analogously for $Y$ and $Z$.

Indeed, in the symmetric quantum realization, the three mixed
correlators can be represented as scalar products of three unit vectors.
Consequently, their correlation `Gram' matrix 
\begin{equation}
G=\left(
\begin{array}{ccc}
1 & X & Y\\
X & 1 & Z\\
Y & Z & 1
\end{array}
\right)
\end{equation}
must be positive semidefinite. The condition $\det G\geq0$ immediately
gives
\begin{equation}
1+2XYZ-X^2-Y^2-Z^2\geq0 \text{ ,}
\end{equation}
see again Figs. \ref{fig:tetrahedron} and \ref{fig:sections} for a comparison of the local and quantum regions.

An exemplificative quantum `twin' system is given by $\hat{\rho}=\left\vert \Psi\right\rangle
\left\langle \Psi\right\vert $ with $\left\vert \Psi\right\rangle $ referring
to a two-fermion system with spin given by
\begin{equation}
\left\vert \Psi\right\rangle =\frac{1}{\sqrt{2}}\left(  \left\vert
++\right\rangle +\left\vert --\right\rangle \right)
\end{equation}
In this maximally entangled case, the twin property can be seen directly from
the corresponding probabilities. For two polarization measurements at
angles $\theta_{q_{1}}$ and $\theta_{q_{2}}$ (thus, at each site the same three analyzer  angles $\theta_1, \theta_2, \theta_3$ are considered), the probability of obtaining the
same result is
\begin{equation}
P_{Q}(1,1|q_{1},q_{2})+P_{Q}(-1,-1|q_{1},q_{2})
=\cos^{2}(\theta_{q_{1}}-\theta_{q_{2}}) .
\end{equation}
Hence, for equal measurement settings, the probability is one (Eq. (\ref{quantumtwin}) is fulfilled).
Thus, although the individual outcomes are probabilistic, the two
particles are exact twins whenever the same question is asked. In this example, the values $(X,Y,Z)$ lie
exactly on the boundary of the elliptope, $1+2XYZ-\left(  X^{2}+Y^{2}%
+Z^{2}\right)  =0$. This limit corresponds to the Tsirelson bound for this example \cite{Tsirelson:1980}.

\begin{figure}[htb]
\centering
\includegraphics[width=0.72\textwidth]{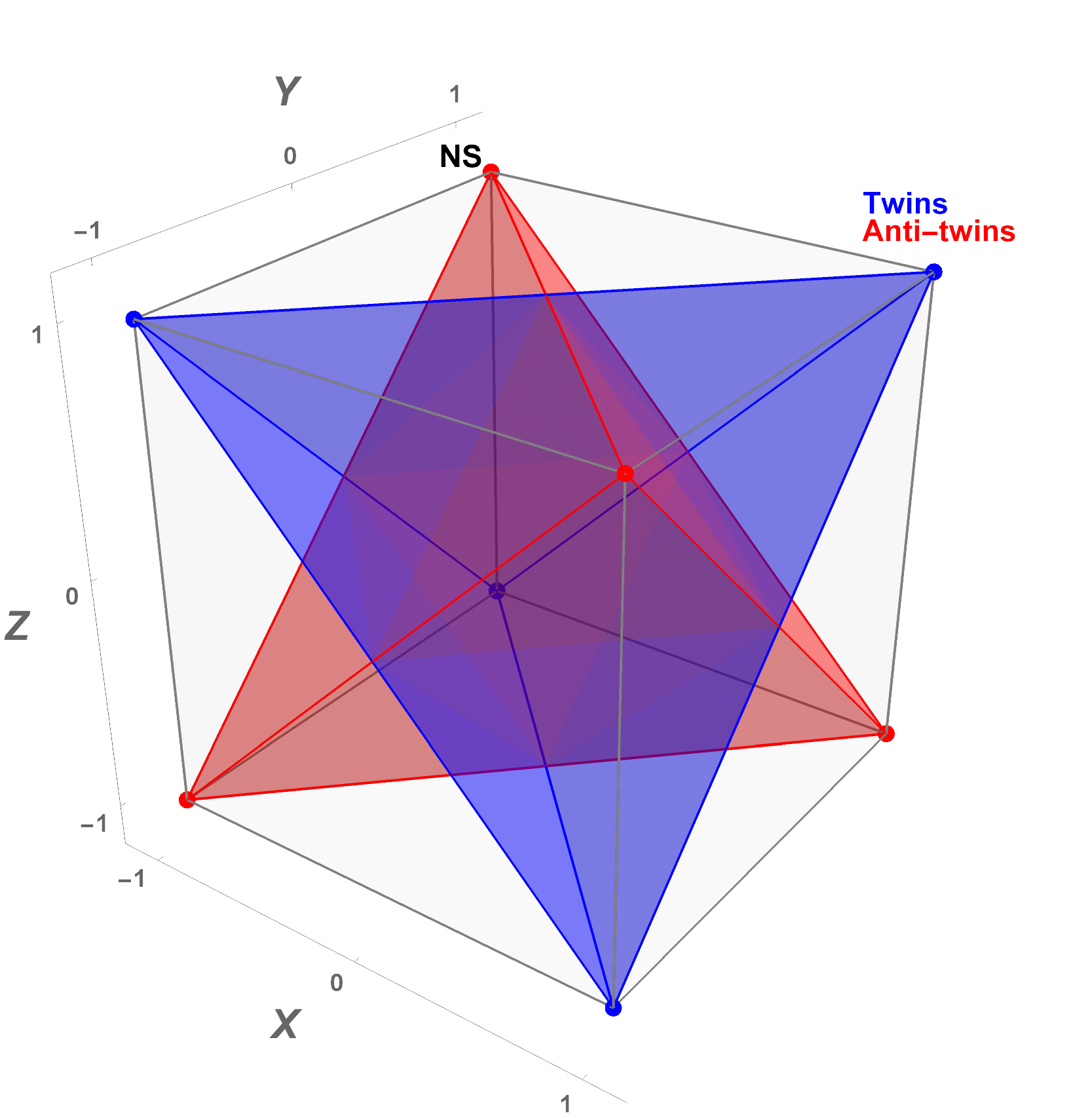}
\caption{The local tetrahedron corresponding to twins (blue) and anti-twins
(red) in the $(X,Y,Z)$ correlation space. The anti-twin tetrahedron is
obtained from the twin tetrahedron by the inversion
$(X,Y,Z)\rightarrow(-X,-Y,-Z)$. The two tetrahedra are therefore related
by reflection through the origin and together occupy complementary sets
of four vertices of the no-signalling cube
$\mathcal{NS}=[-1,1]^3$.}
\label{fig:twins-antitwins}
\end{figure}

The `twin-case' is indeed typical when considering certain decays. e.g. the decay of the Higgs boson into a pair of fermions \cite{Barr:2024djo,Barr:2021zcp}, baryons created in heavy-ion collisions \cite{STAR:2025njp}, and also photons created by a scalar particle \cite{Giacosa:2007bs}. Thus, the tetrahedron construction can be tested by particle experiments; this is an outlook of the present work.

Finally, we mention the anti-twin scenario. This is the case when
Alice and Bob answer oppositely whenever the same question is posed. If Alice liked dinner, Bob did not or vice versa.
The situation amounts to the replacement
\begin{equation}
(X,Y,Z)\rightarrow(-X,-Y,-Z).
\end{equation}
Hence, the anti-twin tetrahedron is defined by the vertices $(-1,-1,-1),$ $(-1,1,1),$
$(1,-1,1),$ $(1,1,-1)$ with inequalities%
\begin{align}
X+Y+Z &  \leq1\text{ , }-X+Y-Z\leq1\\
-X-Y+Z &  \leq1\text{ , }X-Y-Z\leq1
\end{align}
The quantum elliptope reads $1-2XYZ-\left(  X^{2}+Y^{2}+Z^{2}\right)  \geq0$, see Fig. 3.
The surface is realized for spin-singlet state such as:
\begin{equation}
\left\vert \Psi\right\rangle =\frac{1}{\sqrt{2}}\left(  \left\vert
+-\right\rangle -\left\vert -+\right\rangle \right)
\end{equation}
This case applies to e.g. the decay of the para-positronium into two photons \cite{Moskal:2024fpv}. 

In conclusions, we have presented a simple geometric discussion of Bell inequalities in a
symmetric twin scenario with three questions and two possible answers.
Starting from the elementary breakfast--lunch--dinner example, the local
region in the $(X,Y,Z)$ correlation space emerges as a tetrahedron, while
QM enlarges it to an elliptope, both being contained in the
no-signalling cube. The corresponding anti-twin scenario follows by the simple inversion
$(X,Y,Z)\rightarrow(-X,-Y,-Z)$. Applications to examples with particle decay is left an an outlook.

\bigskip 

\textbf{Acknowledgments}
The author thanks M. Gazdzicki for useful discussions. 
This work was supported by the Polish Minister of Science under the ‘Regional Excellence Initiative’ program (project RID/SP/0015/2024/01).

\bigskip

\textbf{AI disclosure:} ChatGPT (OpenAI) was used to assist in discussing and refining the author's ideas, revising the text and language. The scientific ideas and results are those of the author.

\FloatBarrier

\printbibliography

\end{document}